\documentclass{article}
\usepackage{amsmath}
\usepackage{amssymb}
\usepackage{epsfig}
\usepackage{graphicx}

\begin{document}

\title{A K\"{a}hler and quaternion-K\"{a}hler spacetime structure }
\author{R. Vilela Mendes\thanks{%
rvilela.mendes@gmail.com; rvmendes@ciencias.ulisboa.pt;
https://label2.tecnico.ulisboa.pt/vilela/} \\
CEMS.UL, Faculdade de Ci\^{e}ncias, Universidade de Lisboa}
\date{ }
\maketitle

\begin{abstract}
When real Lorentzian spacetime is embedded into a manifold parametrized by
higher division algebras (complex or quaternion with Hermitean metric) and
the representation constraints of their symmetry groups are made compatible,
a set of quantum numbers is generated that is evocative of those of the
standard model of particle physics. This is taken here as a hint that in
spacetime there is a pseudo-K\"{a}hler or pseudo-quaternion-K\"{a}hler
structure, real spacetime being a submanifold that inherits the symmetry
contraints of the larger ambient manifold.
\end{abstract}

\section{Introduction}

As stated in the abstract, this note describes a K\"{a}hler (KM) and
quaternion-K\"{a}hler manifold (QKM) perspective for the structure of
spacetime. This, of course, might be regarded as a pure mathematical
exercise of academic interest. However, there are some hints suggesting that
the embedding of real Lorentzian spacetime into a manifold parametrized by
higher division algebra $\left( \mathbb{A-}\right) $ numbers (complex or
quaternion) might provide, through symmetry considerations, a rationale for
the profusion of quantum numbers in the standard model of particle physics.

There are essentially two distinct ways to embed real Lorentzian spacetime $%
\mathcal{M}_{\mathbb{R}}$ in a higher division algebra manifold $\mathcal{M}%
_{\mathbb{A}}$. With the reference flat metric $G=\left( 1,-1,-1,-1\right) $
one may consider either%
\begin{equation*}
\Lambda ^{T}G\Lambda =G
\end{equation*}%
or%
\begin{equation*}
\Lambda ^{\dagger }G\Lambda =G,
\end{equation*}%
with $\Lambda _{\alpha \beta }\in \mathbb{A}$, for the extended Lorentz
group. In the case $\mathcal{\mathbb{A}=}\mathbb{C}$ one obtains in the
first case $\Lambda \in SL\left( 2,\mathbb{C}\right) \times SL\left( 2,%
\mathbb{C}\right) $, a $12$ parameter group and in the second $\Lambda \in
U\left( 1,3\right) $, a $16$ parameter group.

The first type of complex extension is the one that has been used in the
analytical continuations of the $S$-matrix, interpretation of complex
angular momentum and in most studies of complex gravity. On the other hand,
the second (Hermitean metric) case is the one that insures a uniform
reduction to the usual Lorentz group in each one of the fibers of the
Grassmannian of four dimensional frames in $\mathcal{M}_{\mathbb{A}}$.

The two different extensions of the Lorentz group, complemented with
translations, provide $\mathbb{A-}$extensions of the Poincar\'{e} group with
very different topological properties in $\mathcal{M}_{\mathbb{A}}$. A very
important difference is the role that spinors play. In the first case
spinors in $SL\left( 2,\mathbb{C}\right) $ and $SU\left( 2,2\right) $ \cite%
{Penrose-Book} play a central role in the formulation of the global
properties of $\mathcal{M}_{\mathbb{A}}$ and on the fibration $\mathbb{C}%
^{4}\rightarrow \mathcal{M}_{\mathbb{R}}$. In the second (Hermitean) case,
spinors exist only as elementary representation states in each
four-dimensional frame, but not as global states in $\mathcal{M}_{\mathbb{A}%
} $, reflecting the fact that $U\left( 1,3\right) /SO\left( 1,3\right) $ is
not a spin manifold. Taking this second point of view, the structure of the
representation states of the semidirect product of $U\left( 1,3,\mathbb{A}%
\right) $ with the translations $T\left( 4,\mathbb{A}\right) $ has been
studied in detail in \cite{Vilela-IJMPA} for $\mathbb{A=C},\mathbb{H},%
\mathbb{O}$ \footnote{%
In the octonion case the construction of the representations should be
interpreted in a quasi-algebra framework\cite{Majid}.}. The main conclusions
are:

1. There are no half-integer spin elementary states for the extended Poincar%
\'{e} groups;

2. There are integer spin irreducible representations of the larger groups.
However, when the symmetry groups are still extended to include the discrete
transformations (parity and time inversion), together with positivity of
energy in the real Lorentz submanifolds imply the existence of a
superselection rule between the integer spin states of the larger groups and
those associated to the real Poincar\'{e} groups in the real submanifolds.

The conclusion is that spinor states in each real submanifold cannot be
"rotated" in the larger manifold $\mathcal{M}_{\mathbb{A}}$ away from its
submanifold $\mathcal{M}_{\mathbb{R}}$. They would be confined to the real
submanifolds. On the other hand the integer spin states associated to each
real submanifold can only interact with the global integer spin states by $T$%
-violating interactions.

In the past, the fact that the pseudo-unitary algebras do not possess
elementary spinor representations has led some authors to dismiss them as
spectrum-generating algebras. However the non-existence of spinor states, as
linear irreducible representations, is their most interesting feature. The
nonexistence of spinors as linear representations, does not mean that they
cannot be implemented in a non-linear way. This may be done by expressing
the groups as principal bundles, with structure groups the subgroups of the
real submanifolds, and then constructing the associated vector bundles. This
has been carried out in \cite{Vilela-MPLA} \cite{Vilela-NPB} for massive and
massless states. For example, for massive states the isotropy group is $%
U\left( 3,\mathbb{C}\right) $ for the larger group and $SO\left( 3,\mathbb{R}%
\right) $ for the real submanifold. $U\left( 3,\mathbb{C}\right) $ is looked
at as a principal bundle with base $W=U\left( 3,\mathbb{C}\right) /SO\left(
3,\mathbb{R}\right) $ and fiber $SO\left( 3,\mathbb{R}\right) $. $W$ is not
a spin manifold and to convert it into a spin manifold one has to add to it,
by a Whitney sum, an additional $Sp\left( 1\right) \simeq SU\left( 2\right) $
manifold. In the process new quantum numbers are added to the spinors of the
associated vector bundle. In addition one obtains a degeneracy of states
with identical quantum numbers in the base $W$.

A simple reasoning, that applies whether or not the base of the principal
bundles is a spin manifold, is to require the representations of both the
larger group and the groups of the submanifolds to match each other.
Therefore, when spinors are implemented in the fibers, some additional
spinor quantum numbers have to be added to obtain, by coupling, the integer
spin states of the larger group.

In conclusion: requiring in this way the states in the real submanifold 
\textit{to inherit the symmetries of the larger complex group}, a set of
additional quantum numbers was obtained that is evocative of those of
one-generation standard model. As stated in \cite{Vilela-MPLA} \cite%
{Vilela-NPB} the structure is not exactly the same as the algebra of the
standard model. In particular the color-like degenerate states are not
linear representations of $SU\left( 3\right) $, but independent directions
in the base manifold $W.$

A similar discussion has been carried out for the quaternion case in \cite%
{Vilela-arxiv2025}. The quantum number structure that is obtained is
evocative of a $3-$generations standard model.

Is the quantum number structure of the standard model really a consequence
of the symmetries of a larger space where our real spacetime is embedded? Is
this an actual embedding or a mathematical fiction? Can it ever be decided?
Strangely, so far the best hint comes from the algebraic structure of
standard model (see the Appendix).

So far all the symmetry considerations, discussed before, concerned the
tangent space of spacetime. If one wants to further explore the consequences
of embedding real spacetime in larger division algebra spaces, and in
particular what consequences it might have for cosmological models, further
considerations must be called into play. This will be attempted in the next
section for the complex case and in section $3$ for the quaternion case.

\section{Universe as a local pseudo-K\"{a}hler manifold}

The simpler complex embedding of a four-dimensional Lorentzian spacetime
(with symmetry group $SO\left( 1,3\right) \circledS T\left( 4,\mathbb{R}%
\right) $) is into a flat manifold $\mathcal{M}_{\mathbb{C}}^{0}$ with
symmetry group $U\left( 1,3\right) \circledS T\left( 4,\mathbb{C}\right) $
with coordinates%
\begin{equation}
z^{\mu }=x^{\mu }+iy^{\mu }  \label{2.01}
\end{equation}%
and Hermitean metric%
\begin{equation}
g=\eta _{\overline{\mu }\nu }d\overline{z}^{\mu }dz^{\nu },  \label{2.02}
\end{equation}%
$\left\{ X^{\mu }=\frac{\partial }{\partial x^{\mu }};Y^{\nu }=\frac{%
\partial }{\partial y^{\nu }}\right\} $ being a frame for the tangent bundle 
$T\mathcal{M}_{\mathbb{C}}^{0}$, with multiplication by $i$ a $J$
endomorphism (complex structure)%
\begin{equation*}
JX^{\mu }=Y^{\mu },\hspace{1cm}JY^{\mu }=-X^{\mu }.
\end{equation*}

The metric satisfies%
\begin{equation*}
g\left( JX,JY\right) =g\left( X,Y\right)
\end{equation*}%
$X,Y\in T\mathcal{M}_{\mathbb{C}}^{0}$ and the antisymmetric form%
\begin{equation*}
\Omega (X,Y)=g(X,JY)
\end{equation*}%
is closed, because $J$ is parallel ($\nabla J=0$). Hence $\mathcal{M}_{%
\mathbb{C}}^{0}$ is a pseudo-K\"{a}hler manifold.

All symmetry considerations, here and in Ref.\cite{Vilela-NPB}, assumed real
spacetime as embedded in a larger manifold with local symmetry group the
semidirect product of $U\left( 1,3,\mathbb{C}\right) $ and $T\left( 4,%
\mathbb{C}\right) $, the abelian four complex spacetime translations. This
is the complex Poincar\'{e} group\footnote{%
Also called, by some authors \cite{Barut}, the complex Poincar\'{e} group
with real metric. Hermitean metric would be better, consistent with the K%
\"{a}hler embedding.}. However, like in many other cases \cite{Vilela-JPA} 
\cite{Vilela-SPTP}, it is unlikely that this unstable algebra be an actual
symmetry of Nature. The commutation relations of the Lie algebra of $U\left(
1,3,\mathbb{C}\right) \circledS T\left( 4,\mathbb{C}\right) $ are%
\begin{eqnarray}
\left[ M_{\mu \nu },M_{\rho \sigma }\right] &=&-M_{\mu \sigma }\eta _{\nu
\rho }-M_{\nu \rho }\eta _{\mu \sigma }+M_{\nu \sigma }\eta _{\rho \mu
}+M_{\mu \rho }\eta _{\nu \sigma }  \notag \\
\left[ M_{\mu \nu },N_{\rho \sigma }\right] &=&-N_{\mu \sigma }\eta _{\nu
\rho }+N_{\nu \rho }\eta _{\mu \sigma }+N_{\nu \sigma }\eta _{\rho \mu
}-N_{\mu \rho }\eta _{\nu \sigma }  \notag \\
\left[ N_{\mu \nu },N_{\rho \sigma }\right] &=&M_{\mu \sigma }\eta _{\nu
\rho }+M_{\nu \rho }\eta _{\mu \sigma }+M_{\nu \sigma }\eta _{\rho \mu
}+M_{\mu \rho }\eta _{\nu \sigma }  \notag \\
\left[ M_{\mu \nu },K_{\rho }\right] &=&-\eta _{\nu \rho }K_{\mu }+\eta
_{\mu \rho }K_{\nu }  \notag \\
\left[ M_{\mu \nu },H_{\rho }\right] &=&-\eta _{\nu \rho }H_{\mu }+\eta
_{\mu \rho }H_{\nu }  \notag \\
\left[ N_{\mu \nu },K_{\rho }\right] &=&-\eta _{\nu \rho }H_{\mu }-\eta
_{\mu \rho }H_{\nu }  \notag \\
\left[ N_{\mu \nu },H_{\rho }\right] &=&\eta _{\nu \rho }K_{\mu }+\eta _{\mu
\rho }K_{\nu }  \notag \\
\left[ K_{\mu },H_{\rho }\right] &=&0  \label{2.1}
\end{eqnarray}%
with $\mu ,\nu ,\rho ,\sigma \in \left\{ 0,1,2,3\right\} $, the
antisymmetric $M_{\mu \nu }$ are the generators of real rotations and boosts
in $4$ dimensions with metric $\eta =(1,-1,-1,-1)$, the symmetric $N_{\mu
\nu }$ the generators of complex transformations and $H_{\mu }$ and $K_{\mu
} $ the generators of real and complex translations.

The minimal deformation that stabilizes this algebra is\footnote{%
Reduction (contraction) to the complex Poincar\'{e} group would be obtained
by $\frac{M_{\mu 4}}{R}\rightarrow P\mu $, when $R\rightarrow \infty $}%
\begin{eqnarray}
\left[ M_{ab},M_{cd}\right] &=&-M_{ad}\eta _{5bc}-M_{bc}\eta
_{5ad}+M_{bd}\eta _{5ca}+M_{ac}\eta _{5bd}  \notag \\
\left[ M_{ab},N_{cd}\right] &=&-N_{ad}\eta _{5bc}+N_{bc}\eta
_{5ad}+N_{bd}\eta _{5ca}-N_{ac}\eta _{5bd}  \notag \\
\left[ N_{ab},N_{cd}\right] &=&M_{ad}\eta _{5bc}+M_{bc}\eta
_{5ad}+M_{bd}\eta _{5ca}+M_{ac}\eta _{5bd}  \label{2.2}
\end{eqnarray}%
with $a,b,c,d\in \left\{ 0,1,2,3,4\right\} $ and metric $\eta
_{5}=(1,-1,-1,-1,g_{44})$. With $g_{44}$ either $-1$ or $+1$ these are the
stable algebras of $U\left( 1,4,\mathbb{C}\right) $ or $U\left( 2,3,\mathbb{C%
}\right) $. They are the symmetry algebras of flat complex pseudo-Euclidean
spaces in $1+4$ or $2+3$ dimensions.

To obtain a space $\mathcal{M}_{\mathbb{C}^{4}}$ of four complex dimensions
with the same symmetry, one starts from a complexified flat five dimensional
space $\mathcal{M}_{\mathbb{C}^{5}}$ with metric $\eta _{5}$ which, by the
same considerations as above, is a pseudo-K\"{a}hler space. Then, spaces of
smaller complex dimension with the same symmetry are obtained in
submanifolds defined by invariant homogeneous polynomials. A simple choice
would be to restrict the coordinates by%
\begin{equation}
\left\vert z_{0}\right\vert ^{2}-\left\vert z_{1}\right\vert ^{2}-\left\vert
z_{2}\right\vert ^{2}-\left\vert z_{3}\right\vert ^{2}+g_{44}\left\vert
z_{4}\right\vert ^{2}=-R^{2}.  \label{2.3}
\end{equation}%
Depending on the $\mp $ sign of $g_{44}$, the four-dimensional spacetimes $%
\mathcal{M}_{\mathbb{C}^{4}}$\ that are obtained are complex versions of de
Sitter or anti-de Sitter spaces.

Let, for definiteness, $g_{44}=-1$ and parametrize the coordinates by%
\begin{eqnarray}
z^{0} &=&R\sinh \tau e^{i\psi _{0}}  \notag \\
z^{1} &=&R\cosh \tau \cos \theta _{1}e^{i\psi _{1}}  \notag \\
z^{2} &=&R\cosh \tau \sin \theta _{1}\cos \theta _{2}e^{i\psi _{2}}  \notag
\\
z^{3} &=&R\cosh \tau \sin \theta _{1}\sin \theta _{2}\cos \theta
_{3}e^{i\psi _{3}}  \notag \\
z^{4} &=&R\cosh \tau \sin \theta _{1}\sin \theta _{2}\sin \theta
_{3}e^{i\psi _{3}}.  \label{2.4a}
\end{eqnarray}%
With this choice (\ref{2.3}) is satisfied and, from the flat invariant
length $ds_{5}^{2}=d\overline{z_{0}}dz_{0}-d\overline{z_{1}}dz_{1}-d%
\overline{z_{2}}dz_{2}-d\overline{z_{3}}dz_{3}-d\overline{z_{4}}dz_{4}$ for
the five-dimensional $\eta _{5}$ metric, one obtains for the complex four
dimensional submanifold $\mathcal{M}_{\mathbb{C}^{4}}$%
\begin{eqnarray}
\frac{ds_{4}^{2}}{R^{2}} &=&d\tau ^{2}+\sinh ^{2}\tau d\psi _{0}^{2}-\cosh
^{2}\tau \left( d\theta _{1}^{2}+\sin ^{2}\theta _{1}d\theta _{2}^{2}+\sin
^{2}\theta _{1}\sin ^{2}\theta _{2}d\theta _{3}^{2}\right.  \notag \\
&&\left. +\cos ^{2}\theta _{1}d\psi _{1}^{2}+\sin ^{2}\theta _{1}\cos
^{2}\theta _{2}d\psi _{2}^{2}+\sin ^{2}\theta _{1}\sin ^{2}\theta _{2}d\psi
_{3}^{2}\right) ,  \label{2.5}
\end{eqnarray}%
the factor that multiplies $\cosh ^{2}\tau $ being the measure of a complex $%
3$-sphere, $d\Omega _{3\mathbb{C}}$. In these coordinates the complex
structure $J$ is%
\begin{eqnarray*}
&&\tau \overset{J}{\rightarrow }\tau \\
&&\psi _{i}\overset{J}{\rightarrow }\psi _{i}+\pi /2
\end{eqnarray*}%
and $J\left( T_{z}\mathcal{M}_{\mathbb{C}^{4}}\right) =T_{z}\mathcal{M}_{%
\mathbb{C}^{4}}$ for all $z\in \mathcal{M}_{4}$. Therefore $\mathcal{M}_{%
\mathbb{C}^{4}}$, being a complex submanifold of $\mathcal{M}_{\mathbb{C}%
^{5}}$, is also a pseudo-K\"{a}hler manifold with respect to its induced
structure \cite{Chen}.

\begin{figure}[!htb]
\centering
\includegraphics[width=0.75\textwidth]{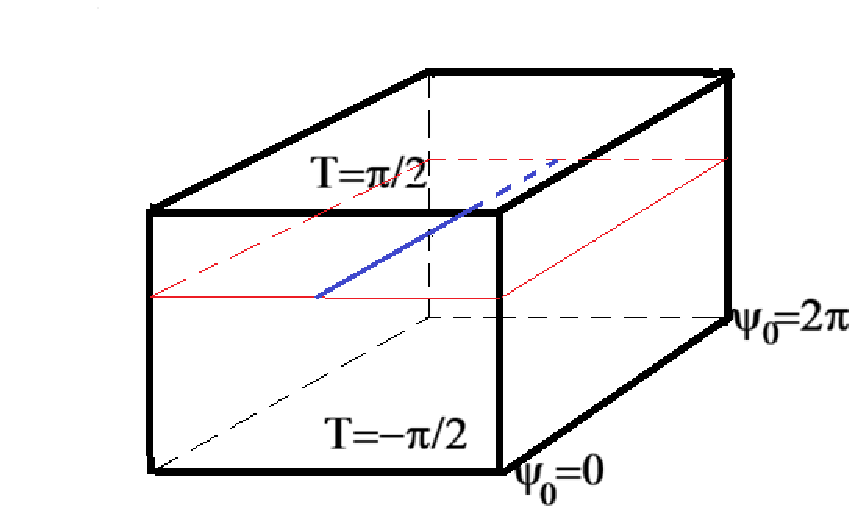}
\caption{Penrose-like diagram for complex de Sitter spacetime. Horizontal
planes (red) are complex 3-spheres and longitudinal lines (blue) complex
2-spheres. Light rays propagate on planes at 45 degrees}
\label{Penrose1}
\end{figure}

$\mathcal{M}_{\mathbb{C}^{4}}$ is a two-time field which, with the change%
\begin{equation*}
\cos T=\frac{1}{\cosh \tau };\hspace{1cm}-\frac{\pi }{2}<T<\frac{\pi }{2},
\end{equation*}%
\begin{equation*}
ds_{4}^{2}=\frac{R^{2}}{\cos ^{2}\tau }\left( dT^{2}+\sin ^{2}Td\psi
_{0}^{2}-d\Omega _{3\mathbb{C}}\right)
\end{equation*}%
may be (conformally) represented by a three-dimensional (Penrose-like)
diagram (Fig.\ref{Penrose1}), each horizontal plane being a complex $3-$%
sphere and each longitudinal line a complex $2-$sphere.

In a real Lorentzian submanifold $\left\{ \psi _{\mu }=0\right\} $, this $%
\mathcal{M}_{\mathbb{C}^{4}}\mathcal{-}$ structure implies a cosmological
constant $\Lambda =\frac{3}{R^{2}}$. In the present symmetry context, a
finite $R$ appears has a natural requirement of stability of the overall
kinematical algebra. A value $R=\infty $ $\left( \Lambda =0\right) $ would
correspond to an improbable exceptional choice of an unstable algebra.
Hence, to look for a mysterious dark energy as the origin of a non-zero $%
\Lambda $ makes as must sense as to look for the reasons for the specific
values of the speed of light or of the Planck constant. We do not know why
these constants have the values they have, we only know that we live in a
generic Universe, rather than an exceptional one with the physical constants
adjusted to unstable algebras.

The strictly local symmetries are well approximated by the semidirect
product $U\left( 1,3,\mathbb{C}\right) \circledS T\left( 4,\mathbb{C}\right) 
$ and stability (genericity) of the symmetry requires $U\left( 1,4,\mathbb{C}%
\right) $. However, as stated in the title of this section, this might also
be only a local structure, other constraints might operate at larger scales.

Complexified forms of spacetime with or without hermitean metrics have been
discussed before by many authors , going back to Einstein \cite{Einstein}.
They appeared either as extensions of the space of solutions or completeness
of geodesics and string propagation, non-local interpretations of Minkowski
or de Sitter space (the twistor approach), etc (\cite{Plebanski}-\cite%
{Witten}). What apparently had not been realized before was the importance
of the distinct nature of the representation spaces for the
pseudo-orthogonal and the pseudo-unitary groups, when they operate as
symmetry groups of the real and the complex spacetimes. And also that when
the symmetries of the complex space are forced upon the real representation
states, new quantum numbers emerge from the bundle Whitney sums.

The non-existence of spinors as representation states when one goes from the
real to the (Hermitean ) complex or quaternionic spacetime has a simple
topological reason. Spinor representations are associated to the spin
groups, which are double covers of orthogonal or pseudo-orthogonal groups, $%
Spin(N)$ being a double cover of $SO(N)$ that allows projective
representations of $SO(N)$ to be lifted to linear representations of $%
Spin(N) $. In contrast, $U(N)$ is simply connected and has no nontrivial
double cover that would give rise to spinor representations.

The complexified de Sitter manifold $\mathcal{M}_{\mathbb{C}^{4}}$ would be
a complex universe without matter, the constant $R$ being simply a constant
of Nature revealing the universe as generic, not an exceptional one ($%
R=\infty $). In this sense it is independent of the matter content of the
Universe. However, to construct cosmological models the presence of matter
should be taken into account. Here also, the representation state
constraints, previously obtained \cite{Vilela-IJMPA}, play a role. Namely,
bulk matter, that is, matter that can be associated to the complex universe
as a whole, is only bosonic matter, fermionic matter being confined to the
Lorentzian four-dimensional submanifolds. In addition, the complex bulk
matter will only interact with the submanifolds' matter (either bosonic or
fermionic) by $T$-violating interactions. This has implications on the
dynamics of the complex space because singular terms (delta-restricted to
the submanifolds) must be represented in the stress-energy tensor. On the
other hand the dynamics on the submanifolds should also take into account
the $T$-violating interactions arising from an eventual distribution of bulk
matter \cite{Vilela-TDark}. This $T$-violating interactions of the bulk
bosonic matter, with the matter associated to the submanifold
representations, might be interpreted as a gravitational interaction, or
might be something different. A simple reasoning, based on the Birkoff
theorem implies that, at large distances from the center, the rotation
velocity of the star galaxies may actually grow with the distance. This
comes from the fact that bulk matter is spread over a six dimensional
spatial volume.

A real four-dimensional Lorentzian spacetime is trivially embedded, as a
submanifold, in $\mathcal{M}_{\mathbb{C}^{4}}$ by $\left\{ \psi _{\mu
}=0\right\} $. A relevant question is what other types of submanifolds can
be identified. Indeed, in a K\"{a}hler manifold there are several typical
interesting families of submanifolds, namely complex, totally real, CR and
slant submanifolds. They are classified by the action of the complex
structure on the tangent bundle. In complex submanifolds $\mathcal{M}$ the
tangent bundle is invariant under the complex structure $J$ ($J\left( T_{z}%
\mathcal{M}\right) =T_{z}\mathcal{M}$). This is the case for $\mathcal{M}_{%
\mathbb{C}^{4}}$ as a submanifold of $\mathcal{M}_{\mathbb{C}^{5}}$. In
totally real manifolds $J\left( T_{z}\mathcal{M}\right) \subset T_{z}^{\bot }%
\mathcal{M}$, that is, $J$ maps tangent spaces into the normal bundle. A
submanifold $\mathcal{N}$ is called a CR-submanifold if there exist a
holomorphic distribution $H=\left\{ H_{z}=T_{z}\mathcal{N}\cap J(T_{z}%
\mathcal{N}):z\in \mathcal{N}\right\} $ and a totally real distribution $%
H^{\perp }$ such that $T\mathcal{N}=H\oplus H^{\perp }$.

Real submanifolds inherit important differential geometry structures from
the K\"{a}hler manifold \cite{Chen2}. With local metric $\eta =(1,-1,-1,-1)$
dimension counting implies that real four-dimensional submanifolds will
necessarily be Lorentzian manifolds. However other types of four-dimensional
submanifolds are possible in the complexified de Sitter spacetime,
four-dimensional universes without time or with two time-like directions.

The extension of spacetime coordinates to higher division algebras \cite%
{Vilela-IJMPA} and the non-linear inheritance of the symmetries of the
larger spaces \cite{Vilela-NPB} \cite{Vilela-arxiv2025} being at the origin
of the quantum numbers of the standard model, would imply that real space
time is embedded in a higher dimensional space, eight-dimensional in the
complex extension case and sixteen-dimensional in the quaternionic case.
Also, the best understood string theory models involve ten-dimensional
spacetimes, with the extra dimensions being curled up on small compact
manifolds. Here instead, no compactification of the extra dimensions is
involved, the relative independence of the four-dimensional submanifolds
being a representation theory effect of the symmetry groups. There are
however possible $T$-violating interactions with some form of bulk bosonic
matter of the extended space.

\section{The quaternion-K\"{a}hler structure}

In a quaternion-K\"{a}hler manifold \cite{Salamon} one has three
anticommuting complex structures. A quaternion-K\"{a}hler manifold may or
may not be a K\"{a}hler manifold, it all depends on the existence of a
global complex structure.

Here the four-dimensional Lorentzian spacetime would be embedded into a
quaternion manifold with strictly local symmetry $U\left( 1,3,\mathbb{H}%
\right) \circledS T\left( 4,\mathbb{H}\right) $ and real dimension $16$. The
restrictions on representation spaces and interactions are similar to the
complex case \cite{Vilela-IJMPA}. Stability of the algebras, that is,
genericity of the symmetries, leads to $U\left( 1,4,\mathbb{H}\right) $
which, with a construction similar to the complex case implies a quaternion-K%
\"{a}hler version of de Sitter space. For the coordinates one may use the
same form as in (\ref{2.4a}) with $e^{i\psi _{\mu }}$ replaced by unit
quaternions.

Could this embedding be physically relevant? As shown in \cite%
{Vilela-arxiv2025} this embedding generates a representation manifold of
elementary states with a structure compatible with the three generations of
the standard model. However the higher dynamical masses of the second and
third generation, suggest that effects of the quaternionic embedding might
only be relevant for higher energies and not, for example, for the galaxy
rotation problem. And could these embeddings be further extended to higher
dimensions? The octonionic case was also discussed in \cite{Vilela-IJMPA},
with the non-associative octonions interpreted in the quasi-algebra sense 
\cite{Majid}. It might suggest a still deeper sub-quark structure.

\section*{Appendix: Quaternionic spacetime and the standard model quantum
numbers}

According to the complex or quaternionic embeddings, massive color states
exist as degrees of freedom in a (symmetric space) $6$ or $18-$dimensional
manifolds ($U\left( 3,\mathbb{C}\right) /SO\left( 3\right) $ or $Sp\left(
3\right) /SO\left( 3\right) $), acted upon by $U\left( 3,\mathbb{C}\right) $
or $Sp\left( 3\right) $ gauge groups.

In the complex case, the base manifold for massive states $U\left( 3,\mathbb{%
C}\right) /SO\left( 3\right) $ the $6$ dimensions are compatible with one
generation of colour quarks and anti-quarks and, in the quaternionic case $%
Sp\left( 3\right) /SO\left( 3\right) $ the $18$ dimensions allow for three
generations of colour quarks and anti-quarks. Both $U\left( 3,\mathbb{C}%
\right) $and $Sp\left( 3\right) $ have no spinors as elementary
representation spaces and if a spin $\frac{1}{2}$ state ($\psi _{1}$) is
attached to the $SO\left( 3\right) $ fibers, extra $SU\left( 2\right) $
bundles (with the same base) should be added, as Whitney sums. A $SU\left(
2\right) $ spinor ($\psi _{2}$) in the additional bundle, coupled with the $%
\psi _{1}$ spinor provides the spin $1$ of the $U\left( 3\right) $ (and $%
Sp\left( 3\right) $) representations, therefore realizing the matching of
the symmetries of the real spacetime ($SO\left( 1,3\right) $) with those of
the embedding complex ($U\left( 1,3,\mathbb{C}\right) $) or quaternionic ($%
U\left( 1,3,\mathbb{H}\right) $) spacetimes. In addition, the extra $%
SU\left( 2\right) $ bundles also endow the spinor states with a pair of
flavour-like quantum numbers.

From the coupling of the two spinors a scalar field is also obtained. Hence
one obtains two Higgs-like states for each generation. Notice that the
requirement of compatibility of the symmetries of the real spacetime with
those of the larger embedding spacetimes only requires the matching of the
representations. In particular the additional $SU\left( 2\right) $ bundles,
might be different for each generation, leading to different mass structures.

For massless states the base manifolds would be $U\left( 2\right) /SO\left(
2\right) $ or $Sp\left( 2\right) /SO\left( 2\right) $ (complex or
quaternionic embeddings) and the matching of the representations also
requires Whitney sums generating flavour-like quantum numbers, but no colour
degeneracy. In the quaternionic case the scalars that are obtained might
again be distinct for each generation, without spoiling the symmetry of the
embedding spacetime, generating different masses for different generations.
The small masses of the neutrinos might have a different origin \cite%
{Vilela-ExtDirac}.

\end{document}